\documentstyle[twocolumn,prb,aps,floats,epsf]{revtex}

\begin{document}
\wideabs{
\title{Interfacial tension behavior of binary and ternary mixtures of 
partially miscible Lennard-Jones fluids: A molecular dynamics simulation.}

\author{Enrique D\'\i az-Herrera}
\address{Departamento de F\'\i sica,
Universidad Aut\'onoma Metropolitana-Iztapalapa,\\
Apartado Postal 55-534, M\'exico 09340, D.F., MEXICO}
\author{Jos\'e Alejandre}
\address{Departamento de Qu\'\i mica,
Universidad Aut\'onoma Metropolitana-Iztapalapa,\\
Apartado Postal 55-534, M\'exico 09340, D.F., MEXICO}
\author{Guillermo Ram\'\i rez-Santiago}
\address{Instituto de F\'\i sica,
Universidad Nacional Aut\'onoma de M\'exico,\\
Apdo. Postal 20-364, M\'exico 01000, D. F., MEXICO}
\author{F. Forstmann}
\address{Institut f\"ur Theoretische Physik,
Freie Universit\"at Berlin, \\
Arnimallee 14, 14195 Berlin, GERMANY}
\maketitle
\vspace{1.cm}
\begin{abstract}
By means of extensive equilibrium molecular dynamics simulations we have 
investigated, the behavior of the interfacial tension $\gamma$ of two immiscible 
symmetrical Lennard-Jones fluids. This quantity is studied as function of 
reduced temperature $T^{*}={ {k_{_B} T}\over \epsilon }$ in the range
$0.6 \leq T^{*} \leq 3.0$. We find that, unlike the monotonic decay
obtained for the liquid-vapor interfacial tension, for the liquid-liquid
interface, $\gamma (T)$ has a maximum at a specific temperature. We also
investigate the effect that surfactant-like particles has on the 
thermodynamic as well as the structural properties of
the liquid-liquid interface. It is found that $\gamma$ decays monotonically 
as the concentration of the surfactant-like particles increases. 
\end{abstract}
}
\section{Introduction}
\label{intro}
The investigation of interfacial tension behavior in complex fluids is of 
relevance from the theoretical as well as from the practical point of view 
\cite{compfluids}. In particular, the nature of the liquid-vapor interface 
has been extensively studied both analytically\cite{anlytlv} and numerically, 
using Monte Carlo\cite{mrao76,mrao79,plinse87,mcslv} and Molecular Dynamics 
\cite{mdslv,li95} algorithms. On the other hand, the properties of the 
liquid-liquid interface, which are of importance in different biological 
systems\cite{biology} as well as in various technological applications 
\cite{technology}, have received  significantly less attention. One reason 
may be the complexity of the topology of the phase diagram for liquid 
mixtures. It is known that the intermolecular properties of a single component 
fluid have no important effect on the topology of its thermodynamic phase 
diagram leading to the concept of corresponding states. By contrast, for 
liquid mixtures the interplay between the molecular 
properties of the components as well as the type of the intermolecular 
potential between different species leads to a highly non-trivial 
topology of the thermodynamic phase diagram\cite{rowlin82,nielaba97}.
Recently, the properties of liquid-liquid planar interfaces have  begun to
be studied by means of molecular dynamics simulations using Lennard-Jones (LJ)
interactions\cite{toxvaerd95,stecki95,scott94,meyer88}, by Monte Carlo 
simulations of benzene-water interface\cite{plinse87} and hexanol-water
\cite{gao88} as well as by density functional theory\cite{stas}. 
In addition, there have also been molecular dynamics simulations for the 
interface of water-alcohols using more complicated intermolecular potentials 
that include the relevant molecular degrees of freedom.
\cite{benjamin92,zhang95,feller95,aldert93,locker90}.
The density and pressure  profiles of a symmetrical binary mixture of LJ 
fluids at the reduced temperature $T^{*}=0.827$ was studied\cite{meyer88} 
in a region of the phase diagram where the system is separated into two 
phases.
It was found that as the system becomes more miscible there is more 
diffusion of particles from one phase to the other and, as a consequence, 
the interfacial tension decreases. Some dynamic properties of this system
were investigated\cite{scott94} at two reduced densities, $\rho^{*}=$0.8 
and 1.37 for several temperatures from which a crude sketch of the phase 
diagram was suggested. A third type of particle representing a simple model 
of amphiphilic molecule was introduced finding that for low concentrations
(of the order of 6\%) there is no important effect on the binary 
fluid interface. In recent papers\cite{toxvaerd95,stecki95} the
structural properties of the above mentioned binary mixture was
investigated at low temperatures and high pressures finding a stable 
oscillatory behavior in the density profiles close to the interface. 
These oscillations decrease as the temperature increases. 
It is also suggested that as pressure grows the interfacial tension increases.
A recent density functional investigation of a binary fluid mixture\cite{stas}
shows that the interfacial tension as a function of temperature can have a 
maximum. 

With the purpose of getting a more complete understanding of the
thermodynamic and structural properties of this liquid-liquid interface
we have performed extensive equilibrium molecular dynamics simulations.
The main results found in this work are: (i) the non-monotonic behavior of 
the interfacial tension as a function of temperature  and (ii) the monotonic
decay of this quantity as the concentration of a third type of particle,
amphiphilic-like, increases. The layout of this paper is as follows: 
In section II we
introduce the model and details of the simulations. In section III we
present the results for the thermodynamic as well as structural properties of
the binary and ternary  systems. Finally, in section IV the conclusions
of this investigation are presented.   

\section{Model and simulations.}
\label{model}
We consider a symmetrical binary mixture of partially miscible Lennard-Jones 
fluids.
The intermolecular interactions $F_{_{AA}},F_{_{BB}}$ and $F_{_{AB}}$ between
particles A-A, B-B and A-B are described by modified LJ potentials that yield
the forces,
\begin{equation}
\label{uab}
F_{_{XY}}(r)=\left\{ \begin{array} {ll} 
 \frac{24}{r}\epsilon \left[ 2 \left( \frac{\sigma}{r} \right)^{12} - 
\alpha_{_{XY}} \left( \frac{\sigma}{r} \right)^6 \right] & \mbox{ if } r \leq 
R_{c} \\
0 & \mbox{ if } r > R_{c} \end{array} \right.
\end{equation}
where $\epsilon$ and $\sigma$ are the same for all interactions
and the parameter $\alpha_{_{XY}}$ controls the 
miscibility of the two fluids. For a partially miscible binary fluid we choose 
$\alpha_{_{AA}} = \alpha_{_{BB}}=1$  and $0 \leq \alpha_{_{AB}} < 1$.
The interaction between particles of the same type is energetically more 
favourable, stronger binding, than the interaction A-B between different 
particles. For this reason one would expect a separation of the species 
at lower temperture when the entropy looses versus potential energy. With 
the aim of understanding the process by which a ``surfactant" weakens the 
interfacial tension and leads to the mixed state, a third species C is 
introduced to try to emulate a surfactant-like particle. In this case we 
consider the same parameters in Eq. \ref{uab} and 
$\alpha_{_{CC}} = \alpha_{_{AC}} = \alpha_{_{BC}} = 1$.

We have carried out extensive equilibrium molecular dynamics (MD) simulations
using the (N,V,T) ensemble for the particular miscibility parameter value
$\alpha_{_{AB}}=0.5$. The range of the inter-molecular potential was set 
equal to three times the particle diameter $\sigma$ unless otherwise stated. 
In most of the simulations a total of N=$1728$ particles were used. Nonetheless,
to check for finite size effects we also carried out simulations 
with N=$2592$ particles. 
These particles are placed in a paralelepiped  of volume
$L_{x}\times L_{y}\times L_{z}$, with $L_{x}=L_{y}$ and $L_{z}=2L_x$,
applying periodic boundary conditions in the $x,y$ and $z$ directions.
The particles were initially placed on the sites of an FCC lattice forming a 
perfect planar interface, that is, all particles of type A are on the left 
side of the box  while those of type B are on the opposite side. In this way 
one can obtain a minimum of two interfaces due to  periodic boundary 
conditions. If one starts with a statistical mixture usually more than two 
demixed regions develop giving rise to more than two interfaces.

The three component system initial configuration was chosen 
from an equilibrated separated binary system picking at random 
$N_{c}/2$ particles from type A and $N_{c}/2$ particles from type B and
replacing them by particles of type C. This way of putting the third species 
in the system makes the total density to remain constant. It is customary to
carry out the simulations using the following reduced units for the distance
$r^*={ {r}\over{\sigma} }$, particle linear momentum 
$p^*={ {p}\over {\sqrt{m\epsilon}} }$ and  time
$t^*={ {t}\over{\sigma} }\sqrt{\epsilon/m}$. 
In these definitions $m$ is the mass of each particle, which is taken to be 
the same for all particles, $\sigma$ is the particle diameter and $\epsilon$ 
is the depth of the LJ potential.
Similarly, one can define reduced thermodynamic quantities as follows:
$T^{*}={ {k_{_B} T}\over \epsilon }$ that represents the reduced temperature 
with $k_{_B}$=Boltzmann's constant and $\rho^*={ {\rho}{\sigma^3} }$ for the 
reduced density, with $\rho=N/V$.
The equations of motion were integrated using a fourth order 
predictor-corrector algorithm with an integration step-size of
$\Delta t^*\le 0.005$, which in standard units is of the order of
$10^{-5}$ nanoseconds in the scale of argon.
The particles initial velocities were assigned from a Boltzmann distribution.
The  equilibration times for most of the simulations were of the order of
$10^{4}$ time steps. Thermodynamic quantities were measured every
50 time-step iterations  up to a total of 5$\times 10^{5}$ to $10^{6}$ 
measurements from which averages were evaluated. This amounts to a
simulation time between 5-10 nanoseconds in the scale of argon.
At the start the reduced homogeneous density in the simulation cell was set 
equal to $\rho^*=0.844$, which is close to density of the triple point of 
argon.
Due to the inhomogeneity which develops in the system around the interface,
the densities of the bulk phases later are slightly higher than this
starting density. The bulk densities are evaluated when the system
has been equilibrated and they depend on the conditions of the 
system and on $T^{*}$.

\section{Result and discussion}
\label{res}
\subsection{Binary mixture}
\label{sub1}
Since our interest is to study the structural and thermodynamic properties of
the interface, we have carried out a set of simulations for a sequence of temperatures
below the critical demixing temperature for two system sizes, namely, $N=1728$
and $N=2592$ particles. We have considered these values of $N$ to
find out about possible finite size effects. All the quantities studied show
qualitatively the same tendency for the two system sizes.
The interfacial behavior is 
investigated by calculating the density profiles, the pressure, and the
interfacial tension at different temperatures. The relevant parameters of the
investigated systems are summarized in Tables \ref{tab-1} and \ref{tab-2} 
where the bulk-values of the density and the pressure are specified. 
To emphasize the separated nature of the system, the values of  
the reduced total density in the bulk A-rich phase $\rho^{\rm bulk\>A}$, 
the reduced density of particles A, $\rho^{\rm bulk\>A}_{_A}$, and particles 
B, $\rho^{\rm bulk\>A}_{_B}$, are given. Due to the symmetry of the 
interactions, the B-rich phase is symmetric to the A-rich phase.

\begin{figure}[b]
\epsfxsize=8cm
\epsfysize=4.5cm
\centerline{\epsffile{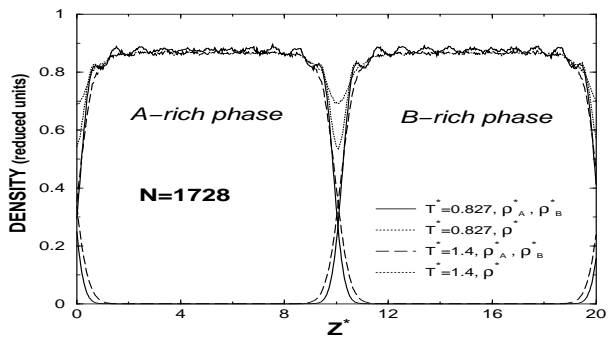}}
\makebox[1cm]{}
\caption{Density profiles for the separated binary mixture with N=1728 particles
and an average reduced total density $\rho^*=0.844$.}
\label{fig-1}
\end{figure}

\begin{figure}[b]
\epsfxsize=8cm
\epsfysize=4.5cm
\centerline{\epsffile{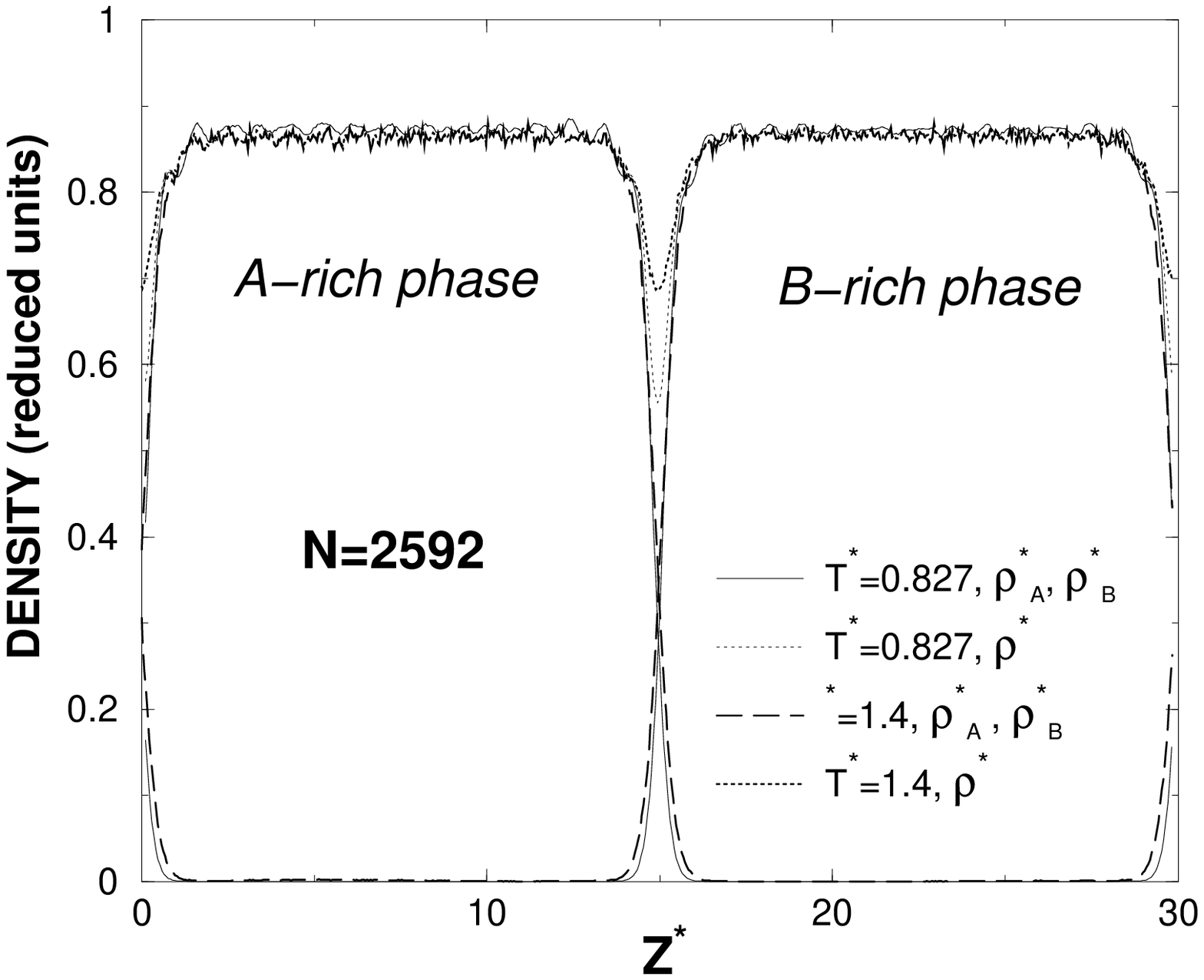}}
\makebox[1cm]{}
\caption{Density profiles for the separated binary mixture with N=2592 particles
and an average reduced total density $\rho^*=0.844$.}
\label{fig-1.1}
\end{figure}

\begin{table}
\begin{tabular}{ccccccc}
Sys. &$T^*$ & $\rho^{\rm bulk}$ & $\rho^{\rm bulk\>A}_{_A}$ &
$\rho^{\rm bulk\>A}_{_B}$ & $P^*_{_n}$ & $\gamma^{*}$\\
\hline
 1 & 0.6   & 0.888 & 0.888 & 0.000 & 0.587  &2.068$\pm$0.012 \\
 2 & 0.827 & 0.876 & 0.876 & 0.000 & 1.964  &2.143$\pm$0.012 \\
 3 & 1.1   & 0.870 & 0.870 & 0.000 & 3.510  &2.186$\pm$0.015 \\
 4 & 1.4   & 0.866 & 0.866 & 0.000 & 5.081  &2.164$\pm$0.018 \\
 5 & 1.7   & 0.863 & 0.863 & 0.000 & 6.555  &2.092$\pm$0.021 \\
 6 & 2.1   & 0.858 & 0.858 & 0.000 & 8.404  &1.941$\pm$0.033 \\
 7 & 2.6   & 0.860 & 0.849 & 0.011 & 10.651 &1.689$\pm$0.039 \\
 8 & 3.0   & 0.859 & 0.842 & 0.017 & 12.368 &1.539$\pm$0.044 \\
\end{tabular}
\vspace{0.5cm}
\caption{Results for the reduced bulk densities and reduced normal pressure
$(P^*_{_n}={{\sigma^3}\over{\epsilon}}P_{_n})$ for the binary mixture with 
$N=1728$ particles.}
\label{tab-1}
\end{table}


\begin{table}
\begin{tabular}{ccccccc}
Sys. &$T^*$ & $\rho^{\rm bulk}$ & $\rho^{\rm bulk\>A}_{_A}$ &
$\rho^{\rm bulk\>A}_{_B}$ & $P^*_{_n}$ & $\gamma^{*}$\\
\hline
 1 & 0.6   & 0.882 (0.7) & 0.882 & 0.000 & 0.458 (22)  &1.986$\pm$0.020 (4) \\
 2 & 0.827 & 0.873 (0.3) & 0.873 & 0.000 & 1.876 (4.5) &2.103$\pm$0.020 (1.9) \\
 3 & 1.1   & 0.869 (0.1) & 0.869 & 0.000 & 3.456 (1.5) &2.180$\pm$0.030 (0.3) \\
 4 & 1.4   & 0.865 (0.1) & 0.865 & 0.000 & 5.068 (0.3) &2.190$\pm$0.050 (1.2) \\
 6 & 2.1   & 0.859 (0.1) & 0.859 & 0.000 & 8.439 (0.4) &1.909$\pm$0.040 (1.7) \\
 8 & 3.0   & 0.860 (0.1) & 0.842 & 0.018 & 12.472(0.8) &1.511$\pm$0.060 (1.8) \\
\end{tabular}
\vspace{0.5cm}
\caption{Results for the reduced bulk densities and reduced normal pressure
for the binary mixture with $N=2592$ particles.
The numbers in parenthesis are the percent differences calculated 
with respect to the corresponding values in table \ref{tab-1}.} 
\label{tab-2}
\end{table}
At sufficiently low temperatures the reduced density of B particles
in the phase of A is very small indicating that the system is fully separated.
As temperature increases the total density of the bulk region  decreases 
slightly because the diffusion of particles through the interface increases, 
making the inhomogeneity smaller and driving the system towards a mixed state.
This can be seen in Figures \ref{fig-1} and \ref{fig-1.1} where the density 
profiles along the $z$ direction (longer side of the box) are shown for 
the  $N=1728$ and  2592 systems, respectively.
The continuous line 
corresponds to a temperature of $T^*=0.827$, which can be considered low,  
(systems 2 in Tables \ref{tab-1} and \ref{tab-2}) while the dashed line 
corresponds to a 
higher temperature $T^*=1.4$ (systems 4 in Tables \ref{tab-1} and \ref{tab-2}).

When looking at the density profiles for the two system sizes studied,
Figures \ref{fig-1} and  \ref{fig-1.1},
we observe that  at $T^{*}=0.827$  these quantities exhibit an oscillatory 
structure at the interfaces. This reminds us about the behavior of the 
density profile in  front of a hard wall, as discussed in reference 
\cite{toxvaerd95}. This kind of structure is also found in the liquid-vapor 
system, but the oscillations are significantly stronger in the liquid-liquid 
interface. As expected, one  also sees from these figures that  the 
oscillations in the bulk region are less pronounced for the 
larger system.

The values of the reduced normal pressure $P^*_{n}$ shown in  
column sixth of Tables \ref{tab-1} and \ref{tab-2} follow an almost 
linear behavior as a function of temperature in the range  
studied. The normal and transversal pressures profiles $P_{n}(z)$, 
$P_t(z)$ were calculated using the definition of the Irving-Kirkwood pressure 
tensor, which for a planar interface is given by \cite{rowlin-capillary,mrao79}
\begin{eqnarray}
\label{eqpnpt}
P_n(z) &=& \rho(z) k_{_B}T\nonumber\\
&-&\frac{1}{2A}\left<\sum_{i\neq j}\frac{z^2_{_{ij}}u'(r_{_{ij}})}
{r_{_{ij}}|z_{_{ij}}|}
\theta\left(\frac{z-z_{_i}}{ z_{_{ij}}}\right)\theta\left(\frac{z_{_j}-z}
{ z_{_{ij}}}\right)\right>,\\
P_t(z) &=& \rho(z) k_{_B}T\nonumber\\
&-&\frac{1}{4A}\left<\sum_{i\neq j}\frac{[x^2_{_{ij}}+y^2_{_{ij}}]
u'(r_{_{ij}})}{r_{_{ij}}|z_{_{ij}}|}
\theta\left(\frac{z-z_{_i}}{ z_{_{ij}}}\right)\theta\left(\frac{z_{_j}-z}
{ z_{_{ij}}}\right)\right>.
\nonumber
\end{eqnarray}

\begin{figure}[t]
\epsfxsize=8cm
\epsfysize=4.5cm
\centerline{\epsffile{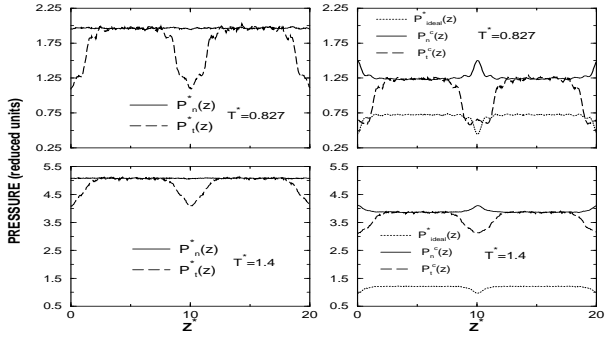}}
\makebox[1cm]{}
\caption{Pressure profiles for the system with $N=1728$ particles.
Left: Normal and transversal reduced pressure profiles
for the separated binary mixture. Right: 
Ideal gas contribution ($P^*_{\rm ideal}=\rho^*T^*$) as well as
contributions from the interaction part
to the transversal ($P^c_t$) and normal ($P^c_n$) components of the pressure 
tensor.}
\label{fig-2}
\end{figure}

In these expressions the first term corresponds to the ideal gas
contribution while the second  comes from the intermolecular forces. 
For a dense system and long ranged intermolecular potentials the latter 
term yields the larger contribution.
On the left side of Figure \ref{fig-2} we show the normal and tangential
components of the pressure tensor as functions of $z$ for $T^*=0.827$ 
and $T^*=1.4$.
The first feature to notice is that $P^*_{n}(z)$ remains constant for both
temperatures as one would expect for a system in thermodynamic
equilibrium. The pressures $P^*_n$ range from 0.25 kbar to 5.18 kbar
in the scale of argon for the systems of Tables \ref{tab-1} and \ref{tab-2}.
On the right side of the same figure the ideal as well as the
configurational parts of $P^*_n(z)$ and $P^*_t(z)$ are plotted. At the lower 
temperature $T^*=0.827$ the pressure profiles clearly show the oscillations 
related to the oscillatory behavior of the density profiles.
One can learn that the configurational contribution is much larger than the 
ideal gas part since we are dealing with a dense system.
These profiles are needed to evaluate the interfacial tension by means of 
the mechanical definition\cite{rowlin-capillary}.

\begin{equation}
\label{ga1}
\gamma = \int^{\rm bulk_B}_{\rm bulk_A} [P_n(z)-P_t(z)]dz.
\end{equation}
Also, a straightforward evaluation of $\gamma$ can be done using the 
Kirkwood-Buff formula\cite{kir-buff,mrao76}
\begin{equation}
\label{ga2}
\gamma = \frac{1}{4A} \left< \sum_{i<j}\left(1-\frac{3z^2_{_{ij}}}
{r^2_{_{ij}}}\right)
r_{_{ij}}u'(r_{_{ij}})\right>.
\end{equation}

In this latter equation there appears an additional factor of $1\over 2$ which comes
from having two interfaces in the system. From the calculational point of
view it is better to use the Kirkwood-Buff formula rather than the mechanical
definition since fluctuations in $P_n(z)$ and $P_t(z)$ may introduce important
inaccuracies in the evaluation of $\gamma$ according to Eq. \ref{ga1}. 
As a matter of fact, we calculated the interfacial tension using both expressions and 
found consistency in the values within the statistics of the simulations.
The range of temperatures, in reduced units, in which $\gamma$ has been 
studied is $0.6 \leq T^{*} \leq 3.0$, that in the scale of argon corresponds to
$72\>K\leq T \leq 359.5\>K$. 

\begin{figure}[t]
\epsfxsize=8cm
\epsfysize=4.5cm
\centerline{\epsffile{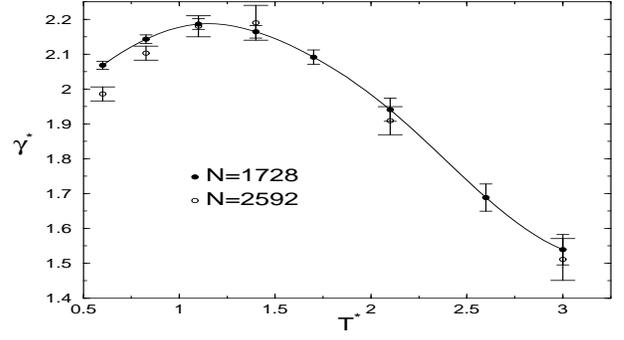}}
\makebox[1cm]{}
\caption{Behavior of the reduced interfacial tension 
($\gamma^*={{\sigma^2}\over{\epsilon}}\gamma$) as a function of
reduced temperature for the systems  with $N=1728$  ($\bullet$)
and $N=2592$  ($\circ$) particles. All lines are a guide to the eye.
The explicit values of $\gamma^*$  together with other thermodynamic 
variables are given in tables \ref{tab-1} and \ref{tab-2}. }
\label{fig-3}
\end{figure}

Unlike for the liquid-vapor interface we find  a non-monotonic 
behavior of $\gamma (T)$ for the liquid-liquid interface. 
In Figure \ref{fig-3} we show the behavior of $\gamma^*$ as a function of 
$T^*$ for systems with $N=1728$ and $N=2592$ particles. 
The main feature of this graph is 
the maximum of the interfacial tension at about $T^{*} \approx 1.1$. 
Such a maximum has been reported recently in the context of density 
functional theory\cite{stas}. It can be understood physically as follows: 
A stronger mixing near the interface introduces weaker A-B bonds
and raises the potential energy. This leads to an increase of $\gamma$,
the free energy of the interface, until at high temperatures the entropy 
contribution drags it down.
We have calculated $\gamma (T)$ for these two systems to chek  for possible
finite size effects.
In doing so  we have to assure that the bulk thermodynamic quantities 
are approximately the same for both systems. In particular, we  monitored very 
closely the bulk density since small variations in this quantity lead to 
significant changes in the pressure of the system at low temperatures. 
So, the important key is to have approximately the same pressure for both 
systems. This can be achieved by adjusting the side $L_{z}$  of the simulational
box for the larger system, mantaining its cross section constant. In this way
we eliminate possible variations of $\gamma (T)$ due to variations in the
cross section area. We found that the optimal average value of $L^{*}_{z}$ 
was 30 for the temperature range studied.  
In  Figure \ref{fig-3} and tables \ref{tab-1} and   \ref{tab-2} 
one also sees that at low temperatures,  the values of $\gamma^{*}(T^{*})$ 
(between the two system  sizes) are at the most 4\% 
different, while at higher  temperatures ($T^{*}> 1.1$) the values are much 
closer to each other.  One also observes that
small variations in the bulk density lead to important changes in the 
pressure at low temperatures. However, at higher temperatures the variations in 
pressure are smaller. This is the expected behavior for a single LJ fluid.
Therefore, differences in $\gamma (T)$ are due to the variations in
pressure, although the former are much smaller than the latter. This 
happens because the interfacial tension is obtained from the difference
between the normal and tangential components of the pressure tensor.
This behavior explains why the differences and consistencies
in the values of $\gamma^{*}(T^{*})$.

\begin{figure}[h]
\epsfxsize=8cm
\epsfysize=4.5cm
\centerline{\epsffile{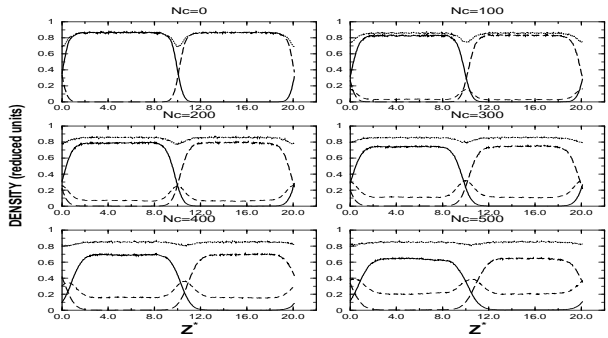}}
\makebox[1cm]{}
\caption{Density profiles for the three component system for different
values of $N_{_C}$ at $T^*=1.4$ and total average  reduced density
$\rho^*=0.844$.
The total number of particles in the system is $N=1728$. Continuous line
correspond to species A, the long dashed to species B, dashed line to
species C, and dotted line to the total density profile.}
\label{fig-4}
\end{figure}


\begin{figure}[b]
\epsfxsize=8cm
\epsfysize=4.5cm
\centerline{\epsffile{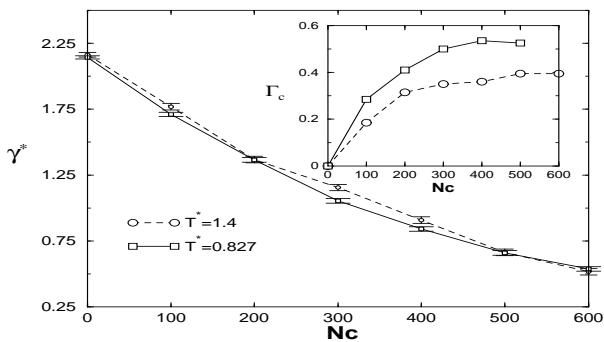}}
\makebox[1cm]{}
\caption{Reduced interfacial tension as a function of $N_{_C}$ for a
system with $N=1728$ particles. The inset shows the excess $\Gamma_{_C}$ 
of particles C at the interface as a function of $N_{_C}$.}
\label{fig-5}
\end{figure}

\begin{table}[t]
\begin{tabular}{ccccc}
 & \multicolumn{4}{c}{$T^*=0.827$} \\ 
\hline \\
Nc & $\rho^{\rm bulk}$ & $\rho^{\rm bulk\>A}_{_A}$ & $\rho^{\rm
bulk\>A}_{_B}$
 & $\rho^{\rm bulk\>A}_{_C}$ \\
\hline
 0   & 0.88 & 0.88 & 0.00 & 0.00 \\
 100 & 0.86 & 0.84 & 0.00 & 0.02 \\
 200 & 0.86 & 0.80 & 0.00 & 0.06 \\
 300 & 0.85 & 0.76 & 0.00 & 0.09 \\
 400 & 0.85 & 0.70 & 0.00 & 0.15 \\
 500 & 0.86 & 0.66 & 0.00 & 0.19 \\
\hline\hline  
 & \multicolumn{4}{c}{$T^*=1.4$}\\
\hline \\
Nc &$\rho^{\rm bulk}$ & $\rho^{\rm bulk\>A}_{_A}$ & $\rho^{\rm
bulk\>A}_{_B}$
& $\rho^{\rm bulk\>A}_{_C}$ \\
\hline
 0   & 0.86 & 0.86 & 0.00 & 0.00 \\
 100 & 0.86 & 0.83 & 0.00 & 0.03 \\
 200 & 0.86 & 0.79 & 0.00 & 0.07 \\
 300 & 0.85 & 0.74 & 0.00 & 0.11 \\
 400 & 0.85 & 0.69 & 0.00 & 0.16 \\
 500 & 0.84 & 0.64 & 0.00 & 0.20 \\
 600 & 0.84 & 0.59 & 0.00 & 0.25 \\
\end{tabular}
\vspace{0.5cm}
\caption{Results for the reduced bulk densities for the ternary
mixture.}
\label{tab-3}
\end{table}

\subsection{Ternary mixture}
\label{sub2}
In this section we investigate the role that a third species plays in 
the interface properties of a demixed binary system. A third species
--- C-particles --- is introduced in the system 
and the behavior of the interfacial tension is studied as a function of 
concentration. The interactions of the 
C-particles $F_{_{CC}}=F_{_{CA}}=F_{_{CB}}$ are all assumed equal to the 
strong interactions $F_{_{AA}}=F_{_{BB}}=F_{_{CC}}$. When placed between the 
A and B, C particles avoid the weak A-B bonds and lower the potential 
energy.
It is found that as $N_{_C}$ increases $\gamma$ decays monotonically. 
This is reminiscent of surfactant-like behavior in some ternary systems. 
In Figure \ref{fig-4} we plot the density profiles of the ternary system as well as 
the reduced total density  for different concentrations of C particles at 
$T^{*}=1.4$. These profiles yield a clear evidence that 
C-particles like to be at the interface position rather than in the bulk 
phases.
In this way the C-particles diminish (screen) the energetically unfavourable
interactions between particles  A-B.  
In Table \ref{tab-3} a summary of the reduced bulk densities of particles 
A, B and C in the A-rich phase for $T^{*}=0.827$ and $T^{*}=1.4$ is given. 
As $N_{_C}$ increases the density of B in the A-rich phase also increases. 
This means that the C particles {\it help} the B particles to diffuse into 
the A-phase because the B particles can be solvated by C and therefore avoid 
the weak A-B interactions. 
As $N_{_C}$ becomes sufficiently large this mechanism drives the system to a 
mixed state. Therefore, the C particles act like a surfactant or emulgator.
This mechanism may have potential technological implications when trying
to design compatibilizers.
In Figure \ref{fig-5} the reduced interfacial tension is plotted as a 
function of $N_{_C}$ for two different temperatures. With increasing bulk 
concentrations of C-particles we see a monotonic decay.
This low concentration behavior is consistent with that found in
reference\cite{smit} where amphiphile like-particles C are introduced.
In such a model $\gamma^*$ as a function of $N_{_C}$ shows a linear
decay in the small amphiphile concentration range. 

In this same figure we also observe that the curve $\gamma^{*}( N_{_C})$ 
at  $T^{*}=1.4$ is  slightly above the curve of $\gamma^{*}( N_{_C})$ at 
$T^{*}=0.827$. 
This is an unexpected result since usually $\gamma^{*}$ decreases when 
$T^{*}$ increases. Nonetheless, we  should recall that these two points lie 
in the region where $\gamma^{*}$ has its maximum as a function of $T^{*}$ 
for the binary system. 
This fact might also be responsible for the shoulder in the curve
for $\gamma^{*}$ at $T^{*}=1.4$ in the region $200<N_{_C}<300$. 
We would like to emphasize that this structure is outside of the 
statistics of the simulations since the error bars are smaller than the 
difference between the points.
We have also evaluated the excess $\Gamma_{_C}$ of C-particles given by 
Equation \ref{excess} and plot this quantity as a function of $N_{_C}$
in the inset of Figure \ref{fig-5}.

\begin{equation}
\label{excess}
\Gamma_{_C} = \int_{\rm{bulk}_{A}}^{\rm{bulk}_{B}} \big(\rho_{_C}(z)-
\rho^{^B}_{_C}\big) \;dz
\end{equation}           

We find that for all values of $N_{_C}$, $\Gamma_{_C}(T^*=0.827)\> > \>
\Gamma_{_C}(T^*=1.4)$ which means that there are more particles of type C 
located at the interface at the lower temperature and therefore they screen 
more effectively the energetically unfavourable interactions between 
A and B particles, diminishing even more the interfacial tension. 
In the range $200 \leq N_{_C} \leq 400$ the excess of C particles
at the interface at $T^{*}=1.4$ increases slower than at $T^{*}=0.827$ 
giving rise to a shoulder in $\gamma^{*}$ as shown in figure \ref{fig-5}.

\begin{figure}[t]
\epsfxsize=8cm
\epsfysize=4.5cm
\centerline{\epsffile{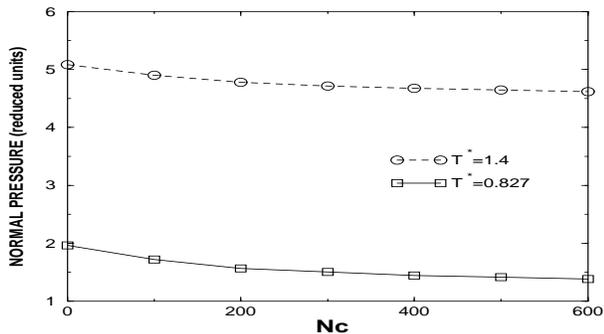}}
\makebox[1cm]{}
\caption{The reduced normal pressure $P^*=P\sigma^3/\epsilon$ as a function
of $N_{_C}$ at two different temperatures for a system with $N=1728$ 
particles.}
\label{fig-6}
\end{figure}

Since in the above results not the pressure but the overall density
$\rho^{*}=0.844$ was kept constant, in Figure \ref{fig-6} we plot the 
behavior of $P^*_{n}$ as a function of $N_{_C}$ for the two temperatures
indicated above. One notes that the hydrostatic pressure decreases 
slowly as $N_{_C}$ increases. This is due to the fact that $N_{_C}/2$ 
particles from fluid A and the same amount from fluid B were switched to 
particles C, replacing the weak A-B attraction by the stronger C-A and C-B 
attraction.
This increases the cohesion and reduces the pressure.
One could be tempted to obtain the upper curve from the lower by
simply adding the ideal gas term $\rho^{*}\Delta T^{*}$,
however, this is not the case because, as noted above with regard to 
Figure \ref{fig-2}, the contribution of the configurational part to the 
pressure is the more relevant one.

\section{Conclusions}
Molecular dynamics simulations for a mixture of two kinds of molecules 
A and B lead to a phase separation with an A-rich and B-rich phase and an 
interface between the phases. This happens if the attractive potential 
between A-B is weaker than that between A-A and B-B and if density and 
temperature are in the two phase region of demixing.  
The interface between the demixed phases has also been investigated. 
We have calculated density profiles at 
different temperatures, the components of the pressure tensor across the 
interface and have evaluated the interfacial tension $\gamma$ for a series of 
temperatures. The important result of the first part of this investigation
is the maximum in $\gamma(T)$ (Figure \ref{fig-3}). 
We explain this result as a consequence of a strong mixing near the interface
that yield weaker A-B bonds thus raising the potential energy, that in turn leads
to an increase in $\gamma(T)$, until at higher temperatures the  interfacial 
free energy decreases due to an increase of the entropy term.

Then we have added a third kind of particle,  ``surfactant-like'', and have 
studied the interfacial tension as a function of its concentration. 
We found that these particles reduce the interfacial 
tension $\gamma$ as its concentration increases. 
This happens because these particles  screen the 
unfavorable A-B interactions and lead to a mixing state of the system.

\section*{Acknowledgments}

We thank S. Iatsevitch and M. Swiderek for helpful discussions.
F.F. would like to thank the Foundation Sandoval Vallarta for financial 
support and the colleagues at the Physics Department of UAM-I for the 
grand hospitality during his visit.
The calculations were carried out on the Power-Challenge computer at the 
UAM-I and at DGSCA-UNAM.
This work is supported by CONACYT research grants Nos. L0080-E and 25298-E.


\begin{thebibliography}{10}
\bibitem{compfluids}
S. A. Safran and N. A. Clark, Editors {\it Physics of Complex and
Supermolecular Fluids}, Wiley New York (1987);
J.M. Charvolin, J.F. Joanny and J. Zinn-Justin, {\it Liquids at Interfaces},
Les Houches Session XLVIII (North-Holland, Amsterdam and New York, 1990).

\bibitem{anlytlv}
C. A. Croxton, {\it Statistical Mechanics of the Liquid Surface}, 
Wiley, New York (1980); J.P. Hansen and I.R. McDonald, {\it
Theory of Simple Liquids}, Academic Press New York (1990).

\bibitem{mrao76}
M. Rao and D. Levesque, J. Chem. Phys. {\bf 65}, 3233 (1976).

\bibitem{mrao79}
M Rao and B. Berne, Mol. Phys., {\bf37}, 455 (1979).

\bibitem{plinse87}
P. Linse, J. Chem. Phys. {\bf 86}, 4177 (1987).

\bibitem{mcslv}
S. W. de Leeuw, C. P. Williams and B. Smit, J. Chem. Phys. {\bf 93}, 2704 
(1990); R. D. Mountain and A. H. Harvey, J. Chem. Phys. {\bf 94}, 2238 (1991).

\bibitem{mdslv}
M. P. Allen and D. J. Tildesley, {\it ``Computer Simulations of Liquids"},
Clarendon Press, Oxford (1987); For a recent review see: C. D. Holcomb,
P. Clnacy and J. A. Zollweg, Mol. Phys. {\bf 78}, 437 (1993) and references 
therein.

\bibitem{li95}
Li-Jen Chen, J. Chem. Phys., {\bf 103}, 10214 (1995).

\bibitem{biology}
A. Alberts, D. Bray, J. Lewis, M. Raff, K. Roberts and J. D. Watsion, in
{\it Molecular Biology of the Cell}, Garland Publications, New York (1989);
R. Lipowsky, Biophys. J. {\bf 64}, 1133 (1993); 
H. G. D\"obereiner, J. K\"as, D. Noppl, I. Sprenger, E. Sackman,
Biophys. J. {\bf 65}, 1396 (1993); F. Julicher and R. Lipowsky,
Phys. Rev. Lett. {\bf 70}, 2964 (1993); 
T. Taniguchi, Phys. Rev. Lett. {\bf 76}, 4444 (1996).

\bibitem{technology}
I. Benjamin,  J. Phys. Chem., {\bf 95}, 6675 (1991);
M. Hayoun, M. Meyer and P. Turq, J. Phys. Chem., {\bf 98}, 6626 (1994).

\bibitem{rowlin82}
J.S. Rowlinson and F.L. Swinton, {\it Liquids and Liquid Mixtures}, 
Butterworth (London 1982).

\bibitem{nielaba97}
N. B. Wilding, F. Schmid and P. Nielaba, Phys. Rev. E {\bf 58}, 2201, (1998).

\bibitem{toxvaerd95}
S. Toxvaerd and J. Stecki,  J. Chem. Phys. {\bf 102}, 7163 (1995).

\bibitem{stecki95}
J. Stecki and S. Toxvaerd, J. Chem. Phys. {\bf 103}, 4352 (1995).

\bibitem{scott94}
W. Scott, F. M\"uller-Plathe and W. F. van Gunsteren,
Mol. Phy. {\bf 82}, 1049 (1994).

\bibitem{meyer88}
M. Meyer, M. Mar\`echal and M. Hayoun, J. Chem. Phys. {\bf 89}, 1067 (1988).

\bibitem{gao88}
J. Gao and W.L. Jorgensen, J. Phys. Chem. {\bf 92}, 5813 (1988).

\bibitem{stas}
S. Iatsevitch and F. Forstmann, J. Chem. Phys. {\bf 107}, 6925 (1997).

\bibitem{benjamin92}
I. Benjamin, J. Chem. Phys., {\bf97}, 1432 (1992).

\bibitem{zhang95}
Y. Zhang, S. E. Feller, B. R. Brooks and R. W. Pastor, J. Chem. Phys., 
{\bf103}, 10252 (1995).

\bibitem{feller95}
S. E. Feller, Y. Zhang and R. Pastor, J. Chem. Phys., 
{\bf 103}, 10267 (1995).

\bibitem{aldert93}
A. R. van Buueren, Siewart-Jan Marrink and H. J. C. Berendsen, 
J. Phys. Chem., {\bf 97}, 9206 (1993).

\bibitem{locker90}
I. L. Carpenter and W. J. Hehre, J. Phys. Chem. {\bf 94}, 531 (1990).

\bibitem{rowlin-capillary}
J. S. Rowlinson and B. Widom, {\it ``Molecular Theory of Capillarity},
Clarendon Oxford, (1982).

\bibitem{kir-buff}
J. G. Kirkwood and F. P. Buff, J. Chem. Phys., {\bf17}, 338 (1949).

\bibitem{smit}
B. Smit, Phys. Rev {\bf A37}, 3431 (1988); B. Smit, A. G. Schlijper,
L. A. M. Rupert, and N. M. van Os, J. Phys. Chem. {\bf 94}, 6933 (1990).

\end{thebibliography}
\end{document}